\documentstyle[qite,hyperref,amsmath,amssymb,graphicx]{article}
\def\Tr{\operatorname{Tr}}

\def\>{\rangle}\def\<{\langle}
\def\sH{\mathcal{H}}
\def\sK{\mathcal{K}}

\def\geq{\geqslant}\def\leq{\leqslant}
\begin{document}

\title{Optimal superbroadcasting of mixed qubit states} \author{
  Francesco Buscemi\makebox[0pt][l]{\addressmark{1}} \\
  buscemi@fisicavolta.unipv.it \and
  Giacomo Mauro D'Ariano\makebox[0pt][l]{\addressmark{1}}\\
  dariano@unipv.it \and
  Chiara Macchiavello\makebox[0pt][l]{\addressmark{1}}\\
  chiara@unipv.it \and
  Paolo Perinotti\makebox[0pt][l]{\addressmark{1}\,\addressmark{2}}\\
  perinotti@fisicavolta.unipv.it }

\seteaddress{\makebox[0pt][r]{\addressmark{1}}%
  QUIT Group, Dipartimento di Fisica ``A. Volta''\\
  \makebox[0pt][r]{\addressmark{2}}%
  INFM-CNR, Unit\`a di Pavia\\
  via Bassi 6, I-27100, Pavia, Italy\\
  TEL: +39-0382-987675 \quad FAX: +39-0382-987563}

\abstract{"Broadcasting", namely distributing information over many users, suffers in-principle
  limitations when the information is quantum. This poses a critical issue in quantum information
  theory, for distributed processing and networked communications. For pure states ideal
  broadcasting coincides with the so-called "quantum cloning", describing an hypothetical ideal
  device capable of producing from a finite number $N$ of copies of a state (drawn from a set) a
  larger number $M>N$ of output copies of the same state. Since such a transformation is not
  isometric, it cannot be achieved by any physical machine for a quantum state drawn from a non
  orthogonal set: this is essentially the content of the "no-cloning" theorem. For mixed states the
  situation is quite different, since from the point of view of each single user a local marginal
  mixed state is indistinguishable from the partial trace of an entangled state, and there are
  infinitely many joint output states that correspond to ideal broadcasting. Indeed, for
  sufficiently large number $N$ of input copies, not only ideal broadcasting of noncommuting mixed
  states is possible, but one can even purify the state in the process. Such state purification with
  an increasing number of copies has been named "superbroadcasting". In this paper we will review
  some recent results on superbroadcasting of qubits, for two different sets of input states,
  corresponding to universally covariant broadcasting and to phase-covariant broadcasting of
  equatorial states.  After illustrating the theoretical derivation of the optimal broadcasting
  channels, we give the maximal purity and the maximal number of output copies $M$ for which
  superbroadcasting is possible. We will see that the possibility of superbroadcasting does not
  increase the available information about the original input state, due to detrimental correlations
  between the local broadcast copies, which do not allow to exploit their statistics. Thus,
  essentially, the superbroadcasting channel simply transfers noise from local states toward
  correlations.  We finally propose a procedure to realize optimal superbroadcasting maps by means
  of optimal pure states cloners.}

\keywords{quantum cloning, quantum broadcasting, Wedderburn
  decomposition, Schur-Weyl duality}

\maketitle

%=====================================================================
\section{Introduction}
%=====================================================================

``Information'' is by its nature {\em broadcastable}. What about when
information is quantum?  Do we need to distribute it among many users?
Indeed, this may be useful in all situations in which quantum
information is required in sharable form, e.~g. in distributed quantum
computation, for quantum shared secrecy, and, generally, in quantum
game-theoretical contexts. However, contrarily to the case of
classical information, which can be distributed at will, broadcasting
quantum information can be done only in a limited fashion. Indeed, for
pure states ideal broadcasting is equivalent to the so-called
``quantum cloning'', which is impossible due to the well-known
``no-cloning'' theorem~\cite{Wootters82,Dieks82,Yuen}. The situation
is more involved when the input states are mixed, since 
broadcasting can be achieved with an output joint state that is
indistinguishable from the tensor product of local mixed states from
the point of view of individual receivers. Therefore, the no-cloning
theorem cannot logically exclude the possibility of ideal broadcasting
for sufficiently mixed states.

In Ref.~\cite{Barnum96} it has been proved that perfect broadcasting is impossible from $N=1$ input
copy to $M=2$ output copies, and for a set of non mutually commuting density operators.  This result
was then considered (see Refs.~\cite{Barnum96} and~\cite{clifton}) as evidence of the general
impossibility of broadcasting mixed states in the more general case in which $N>1$ input copies are
broadcasted to $M>N$ users, for states drawn from a non commuting set. However, in Ref.  \cite{our}
some of the present authors have shown that for sufficiently many input copies $N$ and sufficiently
mixed states the no-broadcasting theorem doesn't generally hold, and it is possible to generate
$M>N$ output local mixed states which are identical to the input ones, and with the input mixed
state drawn from a noncommuting set.  Actually, as proved in Ref.~\cite{our}, it is even possible to
partially purify the local state in the broadcasting process, for sufficiently mixed input state.
Such simultaneous purification and broadcasting was then named ``superbraodcasting''.

The possibility of superbroadcasting does not increase the available information about the original
input state, due to detrimental correlations between the local broadcast copies (see
Ref.~\cite{superbro-pra}), which do not allow to exploit their statistics (a similar phenomenon was
already noticed in Ref.~\cite{Keyl01}). Essentially, the superbroadcasting transfers noise from
local states toward correlations. From the point of view of single users, however, the protocol is a
purification in all respects, and this opens new interesting perspectives in the ability of
distributing quantum information in a noisy environment.

This paper reviews the universal and phase-co\-variant optimal superbroadcasting maps. The two maps
are derived in a unified theoretical framework that is thoroughly presented in Section~\ref{sec1}.
In Sections~\ref{sec2} and~\ref{sec3} we collect the main results concerning universal and
phase-covariant superbroadcasting. In Section~\ref{sec4} we describe a scheme to achieve optimal
superbroadcasting maps of mixed states by means of optimal cloners of pure states.  Finally,
Section~\ref{sec5} discusses the role of correlations in the output states.

%=====================================================================
\section{Symmetric qubits broadcasting}\label{sec1}
%=====================================================================

In deriving optimal maps, we shall extensively use the formalism of
Choi-Jamio\l kowski isomorphism \mbox{\cite{jam,choi}} of CP maps
$\mathcal{E}$ from states on the Hilbert space $\sH$ to states on the
Hilbert space $\sK$, and positive bipartite operators $R$ on
$\sK\otimes\sH$
\begin{equation}\label{eq:choi-jam}
\begin{split}
  &R_{\mathcal E}=\left({\mathcal E}\otimes{\mathcal I}\right)|\Psi^+\>\<\Psi^+|,\\
  &{\mathcal E}(\rho)=\Tr_{\sH}\left[\left(I\otimes\rho^T\right)\
    R_{\mathcal E}\right],
\end{split}
\end{equation}
where $\Psi^+$ is the non normalized maximally entangled state $\sum_m|m\>\otimes|m\>$ in
$\sH\otimes\sH$, and $X^T$ denotes the transposition with respect to the basis $|m\>$ used in the
definition of $\Psi^+$. In term of $R_{\mathcal E}$, the trace-preserving condition for ${\mathcal
  E}$ reads
\begin{equation}\label{trace-pres-abovo}
  \Tr_{\sK}[R_{\mathcal E}]=I_{\sH},
\end{equation}
and covariance under the action of a group $\bf G$ is equivalent to
\begin{equation}
  {\mathcal E}\left(U_g\rho U_g^\dag\right)=V_g{\mathcal E}(\rho)V_g^\dag\Leftrightarrow\left[V_g\otimes U_g^*,R_{\mathcal E}\right]=0,
\end{equation}
where $U_g$ and $V_g$ are the unitary representations of ${\bf G}\ni g$ on the input and output
spaces, respectively, whereas $X^*\doteq (X^\dag)^T$ denotes the complex conjugated of the operator
$X$. In terms of the operator operator $R_{\mathcal E}$ the group-invariance properties from the map
$\mathcal E$ read as follows
\begin{equation}\label{invariance1}
  {\mathcal E}\left(U_g\rho U_g^\dag\right)={\mathcal E}(\rho)\Leftrightarrow\left[I\otimes U_g^*,R_{\mathcal E}\right]=0,
\end{equation}
and
\begin{equation}\label{invariance2}
  {\mathcal E}(\rho)=V_g{\mathcal E}(\rho)V_g^\dag\Leftrightarrow\left[V_g\otimes I,R_{\mathcal E}\right]=0.
\end{equation}

We will consider CP maps $\mathcal B$ from $N$-qubits states to $M$-qubits states, i.~e.
${\sH}=({\mathbb C}^2)^{\otimes N}$ and ${\sK}=({\mathbb C}^2)^{\otimes M}$.  The first requirement
for a broadcasting map is that all receivers get the same reduced state, a requirement that is
achieved by a map whose output is permutation invariant.\footnote{Actually, this is not strictly
  needed, since a joint output state having identical local partial traces is not necessarily
  permutation invariant. However, most figures of merit used for judging broadcasting maps enjoy
  this invariance, in particular the one that we consider in the present paper. Hence permutation
  invariance of the output can be required without loss of generality.} Moreover, there is no loss
of generality in requiring that it is also invariant under permutations of the input copies. This
two simple properties, according to Eqs.~(\ref{invariance1}) and~(\ref{invariance2}), can be recast
as follows
\begin{equation}
[\Pi_\sigma^M\otimes\Pi_\tau^N,R]=0,
\label{perminv}
\end{equation}
where $\Pi_\sigma^M$ and $\Pi_\tau^N$ are representations of the output and input copies
permutations $\sigma$ and $\tau$, respectively. Notice that permutations representations are all
real, whence $\Pi_\sigma^*=\Pi_\sigma$.\par

A useful tool to deal with unitary group representations $U_g$ of a
group $\bf G$ on a Hilbert space $\sH$ is the Wedderburn decomposition
of $\sH$
\begin{equation}
{\sH}\simeq\bigoplus_\mu{\sH}_\mu\otimes{\mathbb C}^{d_\mu},
\end{equation}
where the index $\mu$ labels equivalence classes of irreducible
representations which appear in the decomposition of $U_g$. The spaces
$\sH_\mu$ support the irreducible representations, and
${\mathbb C}^{d_\mu}$ are the multiplicity spaces, with dimension
$d_\mu$ equal to the degeneracy of the $\mu$-th irrep. Correspondingly
the representation $U_g$ decomposes as
\begin{equation}\label{eq:wedder-for-U}
  U_g=\bigoplus_\mu U_g^\mu\otimes I_{d_\mu}.
\end{equation}
By Schur's Lemma, every operator $X$ commuting with the representation
$U_g$ in turn decomposes as
\begin{equation}
  X=\bigoplus_\mu I_{\sH_\mu}\otimes X_{d_\mu}.
\end{equation}

In the case of permutation invariance, the so-called Schur-Weyl
\cite{fulton} duality holds, namely the spaces ${\mathbb C}^{d_\mu}$
for permutations of $M$ qubits coincide with the spaces $\sH_\mu$ for
the representation $U_g^{\otimes M}$ of ${\mathbb SU}(2)$ where $U_g$
is the defining representation. In other words, permutation invariant
operators $Y$ can act non trivially only on the spaces $\sH_\mu$
\begin{equation}\label{eq:schur-Weyl-dual-form}
Y=\bigoplus_\mu Y_\mu\otimes I_{d_\mu}.
\end{equation}
The Clebsch-Gordan series for the defining representation of
${\mathbb SU}(2)$ is well-known in literature
\cite{fulton,messiah,edmonds}, its Wedderburn decomposition being the
following
\begin{equation}\label{eq:wedderburn-for-su2}
  {\sH}\simeq\bigoplus_{j=j_0}^{M/2}{\sH}_j\otimes{\mathbb C}^{d_j},
\end{equation}
where ${\sH}_j={\mathbb C}^{2j+1}$, $j_0$ equals 0 for $M$ even, 1/2
for $M$ odd, and
\begin{equation}
d_j=\frac{2j+1}{M/2+j+1}\binom{M}{M/2-j}.
\end{equation}

Now, the Hilbert space $\sK\otimes\sH$ on which the operator $R$ acts, supports the two permutations
representations corresponding to the output and input qubits permutations, consequently
\begin{equation}
{\sK}\otimes{\sH}\simeq\left(\bigoplus_{j=j_0}^{M/2}{\sH}_j\otimes{\mathbb C}^{d_j}\right)\otimes\left(\bigoplus_{l=l_0}^{N/2}{\sH}_l\otimes{\mathbb C}^{d_l}\right).
\end{equation}
Upon rearranging the factors in the last equation, we can recast the decomposition in a more
suitable way, namely
\begin{equation}
{\sK}\otimes{\sH}\simeq\bigoplus_{j=j_0}^{M/2}\bigoplus_{l=l_0}^{N/2}\left({\sH}_j\otimes{\sH}_l\right)\otimes\left({\mathbb C}^{d_j}\otimes{\mathbb C}^{d_l}\right),
\end{equation}
and to satisfy Eq.~\eqref{perminv} we have the following form for $R$,
according to Eq.~(\ref{eq:schur-Weyl-dual-form}),
\begin{equation}
R=\bigoplus_{j=j_0}^{M/2}\bigoplus_{l=l_0}^{N/2} R_{jl}\otimes\left(I_{d_j}\otimes I_{d_l}\right),
\label{symmbro}
\end{equation}
where $R_{jl}$ acts on ${\sH}_j\otimes{\sH}_l$.  In order to have trace preservation and complete
positivity, the operators $R_{jl}$ must satisfy the constraints
\begin{equation}
  R_{jl}\geq0\,,\quad\Tr_j[R_{jl}]=\frac{I_{2l+1}}{d_j},
\label{cpt}
\end{equation}
where $\Tr_j$ denotes the partial trace over the space ${\sH}_j$. This is the starting point for the
analysis of symmetric qubits cloning devices. Requiring further conditions such as covariance under
representations $V_g^{\otimes N}$, $V_g^{\otimes M}$ of a group $\bf G$, namely
\begin{equation}
\left[V_g^{\otimes M}\otimes {V_g^*}^{\otimes N},R\right]=0,
\end{equation}
will give a further constraint on the operators $R_{jl}$. In the following we will consider the two
cases ${\bf G}={\mathbb SU}(2)$ (universal covariance) and ${\bf G}={\mathbb U}(1)$
(phase-covariance).

Besides Wedderburn decomposition and the related Schur-Weyl duality, another useful tool we will
extensively use is the decomposition of tensor-power states $\rho^{\otimes N}$
\begin{equation}\label{eq:rho-decomposition}
  \rho^{\otimes N}=(r_+r_-)^{N/2}\bigoplus_{j=j_0}^{N/2}\left(\frac{r_+}{r_-}\right)^{J_z^{(j)}}\otimes I_{d_j},
\end{equation}
where $\rho=\frac{1}{2}(I+r\sigma_z)$, and $J_z^{(j)}=\sum_{m=-j}^jm|jm\>\<jm|$ (for a derivation of
identity (\ref{eq:rho-decomposition}) see Refs.~\cite{Cirac99} and~\cite{superbro-pra}). Notice
that the \emph{total} angular momentum component $J_z$ of $N$ qubits is clearly permutation
invariant and can be written as
\begin{equation}
  J_z=\frac12\sum_{k=1}^N\sigma_z^{(k)}=\bigoplus_{j=j_0}^{N/2}J_z^{(j)}\otimes I_{d_j},
\end{equation}
where $\sigma_z^{(k)}$ denotes the operator acting as $\sigma_z$ on the $k$-th qubit, and as the
identity on all remaining qubits.\par

A simple but effective way of judging the quality of single-site output
$\rho'=\Tr_{M-1}\left[{\mathcal B}\left(\rho^{\otimes N}\right)\right]$ is to evaluate the
projection $r^\prime$ of the output Bloch vector over the input one
\begin{equation}
\Tr[\sigma_z\rho^\prime]=r^\prime\,.
\label{singsitra}
\end{equation}
As we will see the single-site output copy $\rho^\prime$ of a covariant broadcasting map commutes
with the input $\rho$, whence $r'$ is indeed the length of the output Bloch vector. The trace in
Eq.~\eqref{singsitra} can be evaluated by considering that the global output state $\Sigma=\mathcal
B(\rho^{\otimes N})$ is by construction invariant under permutations, hence
\begin{equation}
  \Sigma=\bigoplus_{j=j_0}^{M/2}\Sigma_j\otimes I_{d_j},
\end{equation}
and (see Ref.~\cite{superbro-pra})
\begin{equation}\label{eq:rprime}
  r^\prime=\frac 2M\sum_{j=j_0}^{M/2}d_j\Tr[J_z^{(j)}\Sigma_j].
\end{equation}
In the phase-covariant case, according to the usual convention, we more conveniently take $\rho$
diagonal on the $\sigma_x$ eigenstates, and the previous formula is just substituted by
\begin{equation}\label{eq:rprime-phase}
  r^\prime=\frac 2M\sum_{j=j_0}^{M/2}d_j\Tr[J_x^{(j)}\Sigma_j].
\end{equation}
In the following we will use as figure of merit the length $r'$ of the output Bloch vector. This is
actually a linear criterion, which restricts the search of optimal maps among just the extremal
ones. We emphasize that for evaluating broadcasting maps for qubits the length of the output Bloch
vector is a figure of merit more meaningful than the single-site fidelity. Indeed, the case $r'=r$
corresponds to fidelity one, whereas superbroadcasting is achieved for $r'>r$ with fidelity actually
lower than 1.  Moreover, for $r^\prime<r$ maximization of $r'$ still corresponds to maximizing
fidelity, whereas for $r^\prime>r$, one has indeed clones that are purer than the original copies,
which in applications can be more useful than perfect broadcasting---and, moreover perfect
broadcasting can always be achieved by suitably mixing the output states, e.~g. via a depolarizing
channel. We will see that while the map maximizing $r^\prime$ is unique independently of the input
$r$, the perfect broadcasting one (i.~e. with unit single-site fidelity) is not unique and generally
depends on the input purity (the mixing probabilities vary with $r$).  Results will be reported in
terms of the scaling factor for $N$ inputs and $M$ outputs $p^{N,M}(r)\doteq r'/r$, which is usually
referred to as \emph{shrinking factor} or \emph{stretching factor}, depending whether it is smaller
or greater than 1, respectively.

%=====================================================================
\section{Universal covariant broadcasting}\label{sec2}
%=====================================================================

Let us consider now the universal broadcasting, namely the case in
which we impose on the map the further constraint
\begin{equation}
\left[U_g^{\otimes M}\otimes {U_g^*}^{\otimes N},R\right]=0,
\end{equation}
where $U_g$ is the defining representation of the group ${\mathbb
  SU}(2)$. In Ref.~\cite{superbro-pra}, extremal broadcasting maps are
singled out, and the one maximizing the figure of
merit~(\ref{eq:rprime}) is explicitly calculated. The optimal universal map achieves the scaling factor
\begin{equation}
\begin{split}\label{Puniv}
&p^{N,M}(r)=\\&-\frac{M+2}{Mr}(r_+r_-)^{N/2}\sum_{l=l_0}^{N/2}\frac{d_l}{l+1}\sum_{n=-l}^l n\left(\frac{r_-}{r_+}\right)^{n}.
\end{split}
\end{equation}
\begin{figure}
\includegraphics[width=4cm]{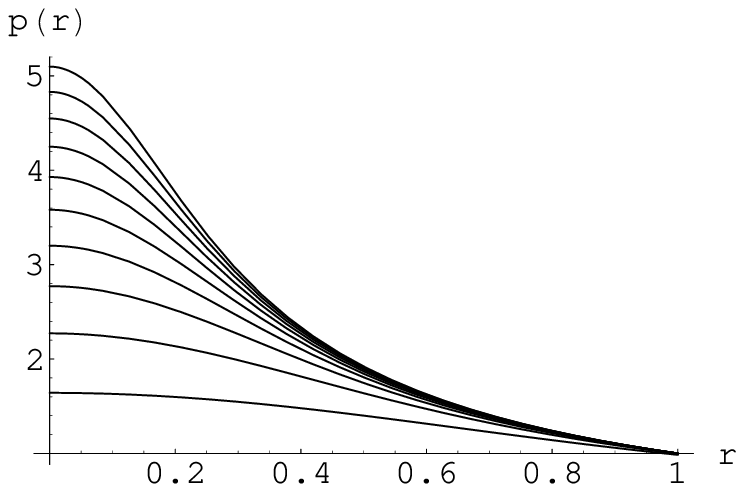}\includegraphics[width=4cm]{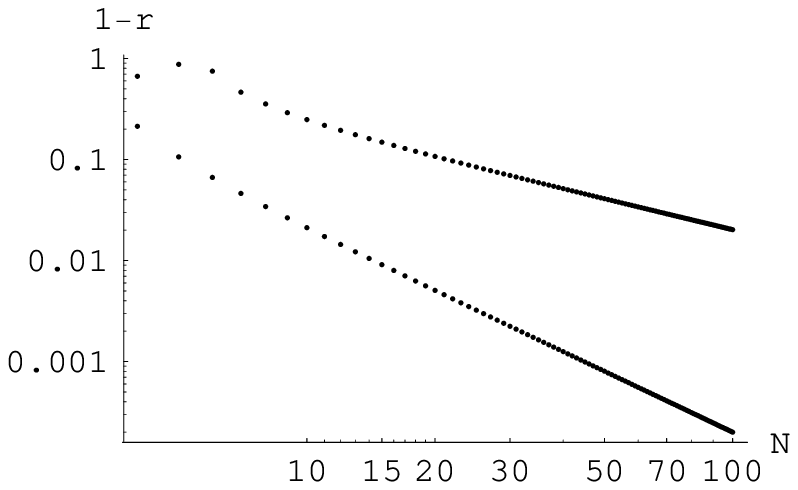}
\caption{Universally covariant broadcasting. {\bf Left:} the behaviour of the optimal scaling
  factor $p^{N,N+1}(r)=r^\prime/r$ in Eq. (\ref{Puniv}) versus $r$, for $N$ ranging from 10 to 100
  in steps of 10. Notice that there is a wide range of values of $r$ such that $p^{N,N+1}(r)>1$,
  corresponding to superbroadcasting. {\bf Right:} logarithmic plot of $1-r_*(N,N+1)$ (lower line) and
  $1-r_*(N,M_*(N))$ (upper line) for $4\leq N\leq100$. The corresponding asymptotic behaviors are 
$2N^{-2}$ and $N^{-1}$, respectively.\label{prplotz}}
\end{figure}
The function $p^{N,N+1}(r)$ is plotted in Fig.~\ref{prplotz} for $N$ from 10 to 100 in steps of 10.
In a range of values of $r$ one has a scaling factor $p^{N,M}(r)>1$, corresponding to
superbroadcasting. This happens for $N\geq4$. The maximum value of $r$ for which superbroadcasting
is possible will be denoted as $r_*(N,M)$ and it is solution of the equation
\begin{equation}
  p^{N,M}(r_*)=1.
\end{equation}
The maximum $M$ for which superbroadcasting is possible for $N$ input copies will be denoted as
$M_*(N)$. It turns out that $M_*(N)=\infty$ for $N>5$, whereas $M_*(4)=7$
and $M_*(5)=21$. The values of $r_*(N,N+1)$ and $r_*(N,M_*(N))$ versus $N$ are also reported in
Fig.~\ref{prplotz}. The corresponding asymptotic behaviors evaluated numerically are $2N^{-2}$ and
$N^{-1}$, respectively.

%=====================================================================
\section{Phase-covariant broadcasting}\label{sec3}
%=====================================================================

Phase-covariant broadcasting corresponds to the constraint
\begin{equation}
  \left[V_\phi^{\otimes M}\otimes {V_\phi^*}^{\otimes N},R\right]=0,
\end{equation}
where $V_\phi=e^{i\phi\sigma_z/2}$ is a representation of the group ${\mathbb U}(1)$ of rotations
along the $z$-axis. Similarly to the case of universal covariance, the optimal map is obtained by
maximizing $r'$ among the extremal maps \cite{superbro-pra}.  The structure of the optimal map
depends only on the parity of $M-N$, similarly to the optimal phase-covariant cloning of pure 
states~\cite{purequbitqutrit,pheconclon}. The scaling factor is given by
\begin{equation}
\begin{split}
  p&^{N,M}_e(r)=\frac{4}{Mr}(r_+r_-)^{N/2}\sum_{l=l_0}^{N/2}d_l\times\\
  &\times\sum_{n=-l}^{l-1}\left[\exp\left(J_x^{(l)}\log\frac{1+r}{1-r}\right)\right]_{n,n+1}\left[J_x^{(j)}\right]_{n,n+1},
\end{split}
\end{equation}
for $M-N$ even, and
\begin{equation}
\begin{split}
p&^{N,M}_o(r)=\frac{4}{Mr}(r_+r_-)^{N/2}\sum_{l=l_0}^{N/2}d_l\times\\
&\times\sum_{n=-l}^{l-1}\left[\exp\left(J_x^{(l)}\log\frac{1+r}{1-r}\right)\right]_{n,n+1}\left[J_x^{(j)}\right]_{n+\frac{1}{2},n+\frac{3}{2}},
\end{split}
\end{equation}
for $N-M$ odd (we use the matrix notation $[X^{(j)}]_{nm}\doteq\<jn|X^{(j)}|jm\>$ for the operator $X^{(j)}$
acting on ${\mathbb C}^{2j+1}$).
\begin{figure}
  \includegraphics[width=4cm]{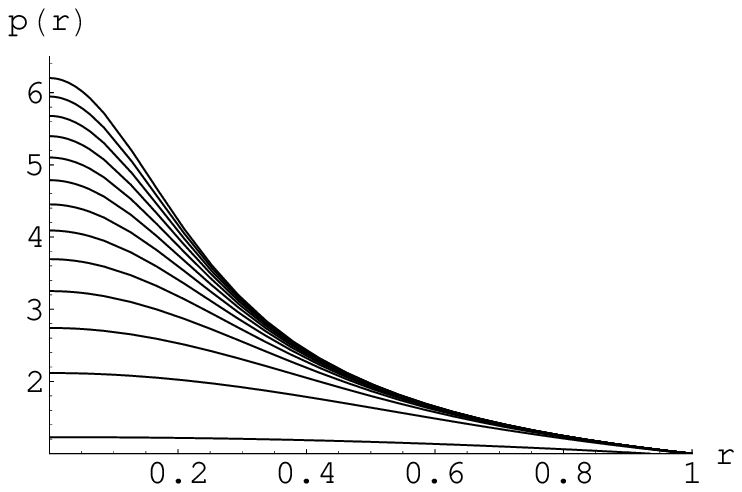}\includegraphics[width=4cm]{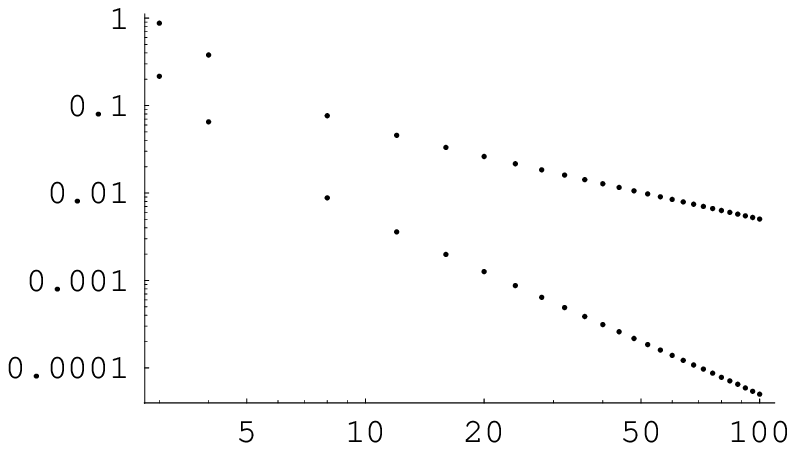}
  \caption{Phase-covariant cloning of equatorial qubits. {\bf Left:} optimal scaling factor
    $p^{N,N+1}(r)=r'/r$ for $N$ ranging from 4 to 100 in steps of 8. {\bf Right:} logarithmic plot
    of $1-r_*(N,N+1)$ (lower line) and $1-r_*(N,M_*(N))$ (upper line) for $3\leq N\leq100$. The
    corresponding asymptotic behaviors are $\frac{2}{3}N^{-2}$ and $\frac{1}{2} N^{-1}$,
    respectively.\label{pr-n+1}}
\end{figure}
The function $p^{N,N+1}(r)$ is plotted in Fig.~\ref{pr-n+1} for $N$ from 4 to 100 in steps of 8. One
has superbroadcasting for $N\ge 3$, with $M_*(3)=12$ and $M_*(N)=\infty$ for $N>3$.\par In
Fig.~\ref{pr-n+1} we also report the plots of the values of $r_*(N,N+1)$ and $r_*(N,M_*(N))$, as for
the universally covariant case, with asymptotic behaviors $\frac{2}{3}N^{-2}$ and $\frac{1}{2}
N^{-1}$, respectively. As expected, the phase-covariant superbroadcaster is always more efficient
than the universally covariant, since the set of broadcasted input states is smaller.

%=====================================================================
\section{Realization scheme}\label{sec4}
%=====================================================================
We propose here a scheme to achieve the optimal $N\to M$ superbroadcasting channels,
for both universal covariance and phase-covariance, using optimal pure state cloners. The method 
exploits a procedure similar to that presented in Ref.~\cite{Cirac99}, based on the
decomposition~(\ref{eq:rho-decomposition}).

The first step is a joint measurement on $\rho^{\otimes N}$ of the observable described by the
orthogonal projectors
\begin{equation}\label{Pja}
  \Pi_{(j,\alpha(j))}=I_{2j+1}\otimes|\alpha(j)\>\<\alpha(j)|,
\end{equation}
where $j_0\le j\le N/2$ labels representation spaces, and $\{|\alpha(j)\>\}$ is an orthonormal basis
spanning the multiplicity space ${\mathbb C}^{d_j}$, $0\le\alpha(j)\le d_j$. For outcome $(l,\chi)$,
the (non normalized) output state after the measurement is
\begin{equation}\label{eq:aftermeas}
  \rho_{(l,\chi)}=(r_+r_-)^{N/2}\left(\frac{r_+}{r_-}\right)^{J_z^{(l)}}\otimes|\chi\>\<\chi|,
\end{equation}
which belongs to the \emph{abstract} subspace ${\mathbb C}^{2l+1}\otimes {\mathbb
  C}^{d_l}\subseteq\left({\mathbb C}^2\right)^{\otimes N}$. By applying a suitable unitary
transformation to the collapsed state (\ref{eq:aftermeas}) it is always possible to rotate it as follows
\begin{equation}\label{eq:afterrot}
\begin{split}
  U_{(l,\chi)}&\rho_{(l,\chi)}U_{(l,\chi)}^\dag=\\
  &(r_+r_-)^{N/2}\left(\frac{r_+}{r_-}\right)^{J_z^{(l)}}\otimes|\Psi^-\>\<\Psi^-|^{\otimes\frac{N-2l}{2}},
\end{split}
\end{equation}
where now the first $2l$ qubits are in the (non normalized) state
$(r_+r_-)^{N/2}\left(\frac{r_+}{r_-}\right)^{J_z^{(l)}}$, whilst the remaining $N-2l$ qubits are
coupled in singlets $|\Psi^-\>$. Finally, once collected the outcome $(l,\chi)$ and rotated the
state to the form~(\ref{eq:afterrot}), one discards the last $(N-2l)$ qubits and applies the
universal (resp.~phase-covariant) optimal $2l\to M$ cloning machine for \emph{pure}
states~\cite{purequbitqutrit,werner} to the remaining $2l$ qubits. One can prove~\cite{unpub} that
using this scheme the optimal $N\to M$ universally covariant (resp.~phase-covariant) broadcasting
map is achieved in average, for universally covariant (resp.~phase-covariant) cloner.
\begin{figure}\begin{center}
\includegraphics[width=6cm]{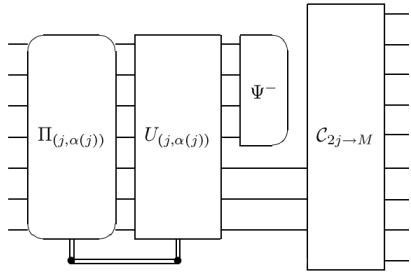}\end{center}
\caption{Sketch of the scheme proposed to realize the optimal $N\to M$ superbroadcaster. 
  On the input state $\rho^{\otimes N}$ the measurement of $\Pi_{(j,\alpha(j))}$ in Eq. (\ref{Pja}) is performed.
  Depending on the measurement outcome, the rotation $U_{j,\alpha(j)}$ is applied to the collapsed
  state. The $(N-2j)$ qubits coupled in singlets are discarded. The remaining $2j$ qubits pass
  through the optimal $2j\to M$ cloner of pure states. At the output we get the $M$ qubits broadcast
  state.}\label{schemasuper}
\end{figure}
The whole procedure is sketched in Fig.~\ref{schemasuper}.

%=====================================================================
\section{Role of correlations}\label{sec5}
%=====================================================================

The optimal superbroadcasting channel allows to obtain a large number of individually good copies of
the same state, starting from fewer---and even more noisy---copies. Indeed, this is possible without
violating the data processing theorem, since the {\em total} amount of information about the
single-site input state $\rho$ is not greater at the output than at the input. The apparently
paradoxical reduction of information on $\rho$ in the presence of purification is due to the fact
that the output copies are not independent, and the total information is not simply the sum of local
contributions. In other words, the phenomenon of superbroadcasting relies on the presence of
correlations at the output, and the superbroadcasting channel can then be regarded as a tool that
moves noise from local states into correlations between them.\par

\begin{figure}
\includegraphics[width=4cm]{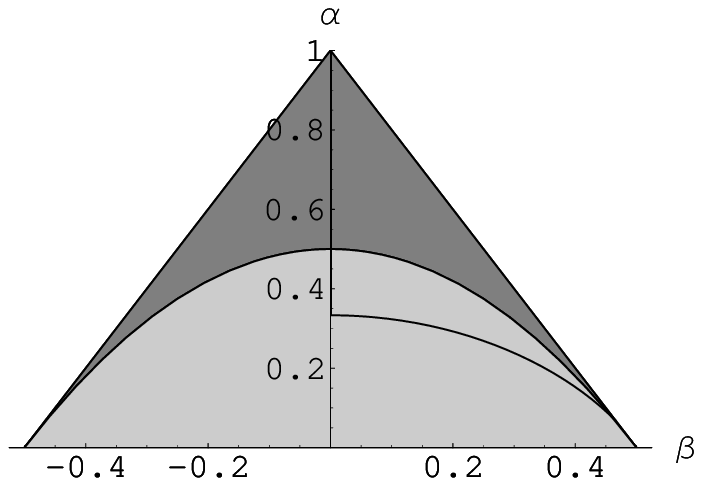}\includegraphics[width=4cm]{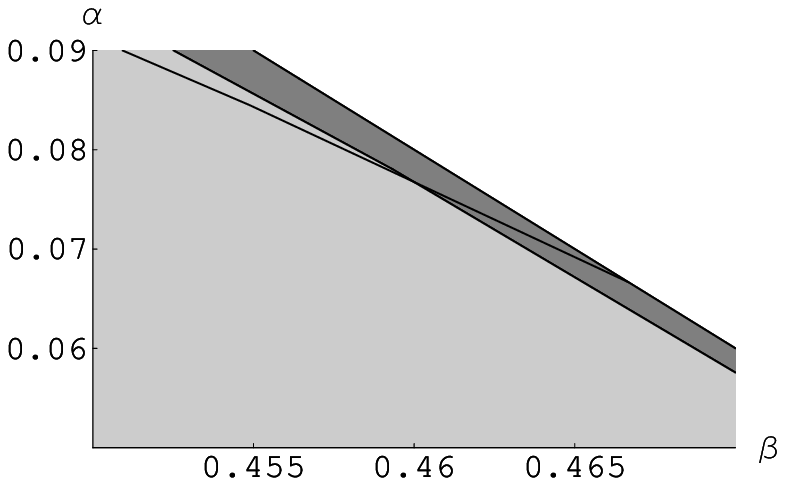}
\caption{Bipartite states $\rho^{(2)}$ at the output of the universally covariant $N\to M$
  superbroadcasting map (the bipartite state corresponds to trace out $M-2$ systems in the global
  output state). {\bf Left:} The triangle in the $\beta,\alpha$ plane contains the symmetric states
  $\rho^{(2)}$ of two qubits commuting with $J^{(1)}_z$ given in Eq. (\ref{symm2}). The light grey
  region represents separable states with $\alpha\leq(1-4\beta^2)/2$, whereas the dark grey region
  represents entangled states. The curve represents the parametric plot $\beta(r),\alpha(r)$ of
  superbroadcast states for $N=4$, $M=5$. {\bf Right:} magnification showing the point in which the
  bipartite output state becomes entangled.\label{concu}}
\end{figure}

\begin{figure}
\includegraphics[width=4cm]{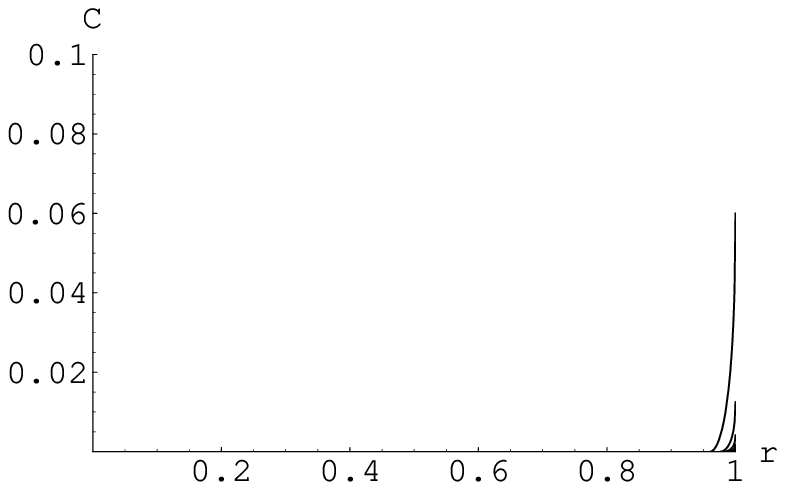}\includegraphics[width=4cm]{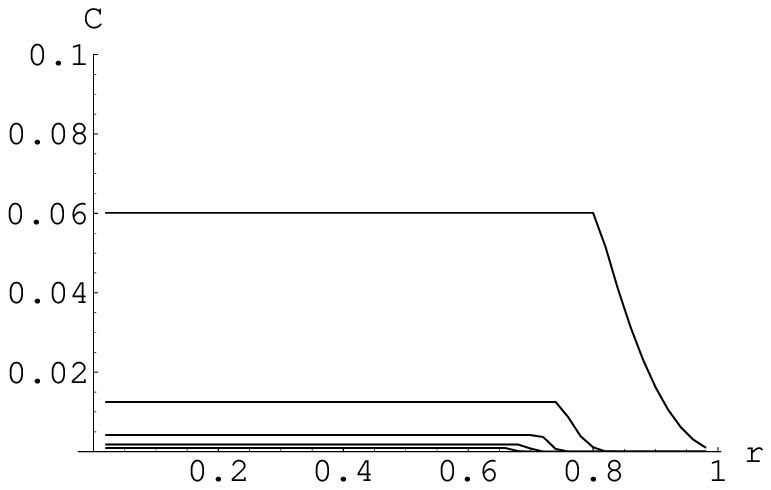}
\caption{Concurrence $C$ versus input purity $r$ of the reduced bipartite states $\rho^{(2)}$ at the
  output of the optimal $N\to N+1$ superbroadcasting channel. {\bf Left:} universally covariant
  case. {\bf   Right:} phase-covariant case. The top curves correspond to $N=2$, while the lower
  plots correspond to even $N$ up to $20$.\label{concuniv}} 
\end{figure}

It is then natural to ask which kind of correlations occur at the output state: are they classical
or quantum? In order to answer this question, we analyzed the bipartite correlations at the output
of the superbroadcasting channels, both for the universally covariant and the phase-covariant cases
(the bipartite state corresponds to trace out $M-2$ systems in the global output state $\Sigma$).
For both types of covariance, the bipartite state is supported in the symmetric subspace of
$({\mathbb C}^2)^{\otimes 2}$ corresponding to the representation $j=1$. In the universally
covariant case, starting from $\rho=\frac{1}{2}(I+r\sigma_z)$ one gets a state commuting with
$J_z^{(1)}$, which can then be parametrized by two real coefficients $\alpha$ and $\beta$ as follows
\begin{equation}\label{symm2}
\rho^{(2)}=\alpha I^{(1)}+\beta J_z^{(1)}+\frac{1-3\alpha}2J_z^{(1)2},
\end{equation}
where $I^{(1)}$ is the projection on the representation $j=1$. The condition for positivity is
simply $0\leq\alpha\leq1-2|\beta|$, corresponding to a triangle in the $\beta,\alpha$ plane, 
with vertices $(-1/2,0)$, $(1/2,0)$ and $(0,1)$. The set of separable states is easily characterized
in terms of $\alpha$ and $\beta$, since the concurrence \cite{wootters} of $\rho^{(2)}$ is nonzero 
iff
\begin{equation}
\alpha\leq\frac{1-4\beta^2}2,
\end{equation}
namely the couple $(\beta,\alpha)$ lies under a parabola which intersects the vertices of the
triangle. In Fig \ref{concu} we plot the states $\rho^{(2)}$ in Eq. (\ref{symm2}) at the output of
the universally covariant $N\to M$ superbroadcasting map.  One can see that mostly the bipartite
correlations are classical, and only for small values of $N$ a certain amount of entanglement is
needed for $r$ approaching purity, whereas for increasing $N$ and $M$ the output states exhibits
essentially classical bipartite correlations. This can be seen also in Fig. \ref{concuniv}, where
the concurrence $C$ is plotted as a function of $r$ for different values of 
$N$, and $M=N+1$. Entanglement is present only for high input purities and low $N$, and vanishes for
increasing $N$.\par
The analysis of correlations in the phase-covariant case does not provide an easy geometrical
visualization, and some insight is given only by the plots of concurrence as a function of $r$,
shown in Fig.  \ref{concuniv}. Also in this case quantum correlations are very small and vanish for
increasing $N$. However, contrarily to the universally covariant case, here concurrence decreases
for $r$ approaching 1.\par

The above results seem to indicate that quantum correlations---at least the bipartite ones---do not
play a crucial role in superbroadcasting.  The natural question is then whether superbroadcasting is
a semi-classical or truly quantum in nature, namely if it can be achieved by measurement and
re-preparation, or if it has a nonvanishing quantum capacity. There are two clues supporting the
hypothesis that the map is truly quantum. The first is that by measurement and re-preparation it is
possible to achieve superbroadcasting, but only sub-optimally, and with scaling factor independent on
$M$ (equal to the optimal factor in the limit $M\to\infty$) \cite{semicla}. The second is that the last
stage of the scheme of the optimal superbroadcasting channel is an optimal pure-state cloner, which
is a purely quantum process \cite{buzhill,gismass,darmac}.

%=====================================================================
%\section{References}
%=====================================================================

\end{document}